\documentclass[twocolumn]{aa}
\usepackage{graphicx}
\usepackage{color}
\usepackage{times}
\usepackage{fancyhdr}
\usepackage{float}
\usepackage{multirow}
\usepackage{amssymb}
\usepackage{natbib}
\bibpunct{(}{)}{;}{a}{}{,} % to follow the A&A style

\newcommand{\tunit}{$\;\textrm{Myr}^{-1}\,\textrm{Mpc}^{-3}$}
\newcommand{\ui}{$\pm$} 
\newcommand{\minks}{80~\%}

\newcommand{\ratemin}{1.1}

\newcommand{\ratere}{7.8}

\newcommand{\ratemax}{11.3}

\begin{document}

\title{Estimation of compact binary coalescense rates from short gamma-ray burst redshift measurements}
\author{Alexander Dietz}
\institute{LAPP, Universit\'{e} de Savoie, CNRS/IN2P3,\\
Chemin de Bellevue, BP 110,\\ 74941 Annecy-le-Vieux CEDEX, France\\ \email{dietz@lapp.in2p3.fr}}

\abstract{}
{We aim to estimate the rate of compact binaries and compare these with similar estimates based on either the observation of these  binaries or population synthesis models. Since the merger of these compact objects are likely to produce short gamma-ray bursts, conclusions about the opening angle of these bursts can be made.}
{We use a set of observed redshift measurements of short gamma-ray bursts to model the rate of these merger events in the nearby universe.
Various corrections are included in the calculation, such as the field-of-view of the satellite missions, the beaming factors of gamma-ray burst and other parameters.}
{The computed rate estimates are compared to other rate estimates, based on observations of binary neutron stars and population synthesis models.
Given the upper limit established by LIGO/Virgo measurements, it is possible to draw conclusions about the beaming angle of gamma-ray bursts.}{}


\maketitle
\titlerunning{Coalescense rates from short GRB redshifts}

%%%%%%%%%%%%%%%%%%%%%%%%%%%%%%%%%%%%%%%%%%%%%%%%%%%%%%%%%%%%%%%%%%%%%%%%%%
\section{Introduction}

Short gamma-ray bursts (GRB) are powerful explosions in the depth of the universe, probably created by the merger of two compact objects, such as two neutron stars (NS-NS) or a neutron star and a black hole (NS-BH). 
These events will be sources of strong gravitational waves (GW), which are being searched for with ground-based GW detectors such as LIGO and Virgo. 
Efforts have been made to estimate the local rates of these events based on the observations of a few known neutron-star binary systems and star population models. 
If short GRBs are indeed created by a merger of two compact objects, the observation of the redshift distribution can be used to determine the local merger rate in a complementary way, which is the main subject of this work. 
Furthermore, known limits on the rate of these events from LIGO/Virgo observations can be used to constrain parameters related to GRB physics, such as the opening angle of the outflows in GRB. 

This paper is organized as follows. After the introduction, the available redshift data for short GRBs is reviewed, before the fit model is explained and the sources of uncertainties are discussed. 
The results of the calculations are then discussed, and compared with other rate estimates and the implications for GRB parameters. 

\subsection{Gamma-ray bursts}

Gamma ray bursts are intensive bursts of high-energy gamma rays, distributed uniformly over the sky, lasting milliseconds to hundreds of seconds. Several thousands bursts have been discovered to date, with the very prominent feature of a bimodal distribution of the burst durations, with a minimum around two seconds \citep{Kouveliotou:1993,Horvath:2002}. 
Bursts with a duration shorter than two seconds are called short GRB's, and bursts lasting longer than two seconds are labeled long GRBs. Long GRBs have been associated with star-forming galaxies and core-collapse supernovae \citep{Campana:2006,Lee:2004xi,Hjorth:2003,Fruchter:2006,Woosley:2006}. 
Short GRBs, on the other hand, had been a mystery for a long time, but early observations with SWIFT  \citep{swift1,swift2} indicated that the progenitor of these events is the merger of two compact objects \citep{eichler:1989, narayan:1992}. 

Besides the merger scenario, some of the short GRBs are probably caused by soft gamma repeaters (SGR), rapidly rotating magnetically-powered neutron stars, creating sporadic 'star quakes' in the crust that generate bursts of gamma radiation \citep{Mereghetti:2008je, woods:2004}. It has been estimated that up to $\sim$25~\% of all SGRBs are caused by SGRs \citep{Tanvir:2005,Levan:2008}.

Throughout this paper, we will use a new GRB classification scheme suggested by Zhang \textit{et.al.} \citep{0004-637X-703-2-1696}, which labels a GRB with a probable merger progenitor as type-I GRB (i.e. the 'short'-GRB class) and GRBs that might have been created by a core collapse supernovae as type-II GRB (i.e. the 'long'-GRB class).  
Although this classification is not unambiguous (there are unclear cases such as GRB~060614, or a sub-class might exist related to e.g. soft gamma-repeaters \citep{Chapman:2008zx,Czerny:2010bd}), this notation is based on the probable underlying physics instead of a single observational quantity, the duration of the burst. 

\begin{table*}[!t!] 
\caption{28 type-I gamma ray bursts with possible redshift measurements and $T_{90}$ duration equal to or shorter than four seconds\tablefootmark{a}.}
\centering
\begin{footnotesize}
\begin{tabular}{|l|l|l|l|l|l|}
\hline
 \textbf{GRB}    & \textbf{instrument}  & \textbf{Optical afterglow} & \textbf{duration [s]}& \textbf{redshift} & \textbf{classification} \\ \hline
020531  & HETE      & yes & 1.0           & 1.0              & probable \\ \hline
\textbf{040924}  & \textbf{HETE}      & \textbf{yes} & \textbf{$\sim$ 1.5}       & \textbf{0.859}            & \textbf{reliable} \\ \hline 
\textbf{050416}  & \textbf{SWIFT/XRT} & \textbf{yes} & \textbf{2.4$\pm$0.2}   & \textbf{0.6535$\pm$0.002} & \textbf{reliable} \\ \hline 
\textit{050509B} & \textit{SWIFT/XRT} & \textit{no?} & \textit{0.03} & \textit{0.2248\ui0.0002}  & \textit{implausible} \\ \hline 
\textbf{050709}  & \textbf{HETE}      & \textbf{yes} & \textbf{0.22$\pm$0.05} &\textbf{0.1606\ui0.0001}  & \textbf{reliable} \\ \hline
\textbf{050724}  & \textbf{SWIFT/XRT} & \textbf{yes} & \textbf{3.0$\pm$1.0}       & \textbf{0.2576$\pm$0.0004}& \textbf{reliable} \\ \hline
050813  & SWIFT/XRT & yes?& 0.6$\pm$0.1   & 0.722            & probable \\ \hline
\textit{050906}  & \textit{SWIFT/BAT} & \textit{no}  & \textit{0.128\ui0.016} & \textit{0.43}             & \textit{implausible} \\ \hline 
\textbf{051016B} & \textbf{SWIFT/XRT} & \textbf{yes} & \textbf{4.0$\pm$0.1}     & \textbf{0.9364}           & \textbf{reliable} \\ \hline
\textbf{051221A} & \textbf{SWIFT/XRT} & \textbf{yes} & \textbf{1.4$\pm$0.2}   & \textbf{0.5465}           & \textbf{reliable} \\ \hline
\textbf{060502B} & \textbf{SWIFT/XRT} & \textbf{no}  & \textbf{0.09\ui0.02}   & \textbf{0.287}            & \textbf{reliable} \\ \hline
060505  & SWIFT/XRT & yes & 4$\pm$1       & 0.089            & probable \\ \hline
\textbf{060801}  & \textbf{SWIFT/XRT} & \textbf{no}  & \textbf{0.5\ui0.1}     & \textbf{1.131}           & \textbf{reliable} \\ \hline 
\textbf{061006}  & \textbf{SWIFT/XRT} & \textbf{yes} & \textbf{130\ui10}      & \textbf{0.4377\ui0.0002}  & \textbf{reliable}\\ \hline 
\textbf{061201}  & \textbf{SWIFT/XRT} & \textbf{yes} & \textbf{0.8$\pm$0.1 }  & \textbf{0.111}     & \textbf{reliable} \\ \hline
\textit{061210}  & \textit{SWIFT/XRT} & \textit{no}  & \textit{85$\pm$5}      & \textit{0.41}             & \textit{implausible} \\ \hline 
061217  & SWIFT/XRT & no  & 0.3$\pm$0.05  & 0.827            & probable \\ \hline
\textit{070209}  & \textit{SWIFT/BAT} & \textit{no}  & \textit{0.10$\pm$0.02}  & \textit{0.314}            & \textit{implausible} \\ \hline
\textit{070406}  & \textit{SWIFT/BAT} & \textit{no}  & \textit{0.7$\pm$0.2}   & \textit{0.11}             & \textit{implausible} \\ \hline
\textbf{070429B} & \textbf{SWIFT/XRT} &\textbf{ yes?}& \textbf{0.5$\pm$0.1}   & \textbf{0.9023\ui0.0002}            & \textbf{reliable} \\ \hline
\textbf{070714B} & \textbf{SWIFT/XRT} & \textbf{yes} & \textbf{64$\pm$5}      & \textbf{0.9225\ui0.0001}  & \textbf{reliable} \\ \hline
070724  & SWIFT/XRT & no  & 0.4$\pm$0.04  & 0.457            & probable \\ \hline 
070810B & SWIFT/XRT & no? & 0.08$\pm$0.01 & 0.49             & probable \\ \hline 
\textbf{071227}  & \textbf{SWIFT/XRT} & \textbf{yes} & \textbf{1.8$\pm$0.4}   & \textbf{0.384}            & \textbf{reliable} \\ \hline
080121  & SWIFT/XRT & no  & 0.7$\pm$0.2   & 0.046            & probable \\ \hline  
\textbf{080520}  & \textbf{SWIFT/XRT} & \textbf{yes} & \textbf{2.8$\pm$0.7 }  & \textbf{1.545}            & \textbf{reliable} \\ \hline  
\textit{090417A} & \textit{SWIFT/BAT} & \textit{no}  & \textit{0.072$\pm$0.018}& \textit{0.088}           & \textit{implausible} \\ \hline
\textbf{090510}  & \textbf{SWIFT/XRT} & \textbf{yes} &\textbf{ 0.3$\pm$0.1}   & \textbf{0.903$\pm$0.003}  & \textbf{reliable} \\ \hline
\end{tabular}
\tablefoot{ The data for these GRB's are taken from \citep{O'Shaughnessy:2007fb,Sakamoto:2007} and from diverse GCN circulars (\texttt{http://gcn.gsfc.nasa.gov/gcn3\_archive.html}) . Reliable redshift values are marked in \textbf{bold}, while implausible redshift values are marked in \textit{italics}.
\tablefoottext{a}{The duration refers to the earth frame. No cosmological correction was applied because of the a priori unknown redshift. }}
\end{footnotesize}
\label{tab:set1}
\end{table*}

\subsection{Associated gravitational wave observations}

As mentioned just above, type-I GRBs might be sources of gravitational waves searched for with gravitational wave detectors, such as LIGO and Virgo. 
The LIGO detectors, described in detail in \citep{LIGOS1instpaper,Barish:1999}, consist of two instruments at two sites in the US, while Virgo consists of one instrument located near Pisa in Italy \citep{0264-9381-23-19-S01}.

Several results of searches for merger signals have been published, with upper limits on the rates of merger events \citep{LIGO03,LIGOS2iul,LIGOS2macho,LIGOS2bbh,DAbbott:2007c}, along with triggered searches for merger signals associated with type-I GRBs \citep{grb070201,S5GRB}, but so far, no GWs have been detected.
These detectors are currently undergoing significant enhancements, before a new data-taking campaign commences in $\sim$2015 with a 10-fold increase in sensitivity with respect to the initial configuration, probing a 1000-fold volume in space.

%%%%%%%%%%%%%%%%%%%%%%%%%%%%%%%%%%%%%%%%
\section{Used data and redshifts}
\label{sec:fitresults}

Because the redshifts are a very crucial piece of information in this work, we re-examine the list of type-I GRBs associated with a redshift measurement in detail, to determine whether the association is justified. 
Table \ref{tab:set1} lists all 28 type-I GRBs with a possible associated redshift over the five year period that we consider; the values are taken from \citep{O'Shaughnessy:2007fb,Sakamoto:2007} and diverse GCN circulars\footnote{\texttt{http://gcn.gsfc.nasa.gov/gcn3\_archive.html}}. 
Each redshift association is classified into three groups: 

\begin{itemize}
 \item \textbf{Reliable redshifts:} GRBs with a very high probability that the associated redshift is correct, which is the case when an optical counterpart has been identified, or when only one galaxy is located in the error-circle of the observation. These GRBs are used in datasets labeled 'A' and 'B'. 

 \item \textbf{Probable redshifts:} GRBs with a good chance that the redshift association is valid, in the case of e.g. two galaxies remaining in the error-circle, or when the observations lead to controversial results. These GRBs are used in the dataset labeled  'B'. 

 \item \textbf{Implausible redshifts:} GRBs with a very low probability that the assigned redshift is correct, as in the cases without identification of an optical counterpart or when many galaxies reside in the error circle. These GRBs are not used in this analysis.
\end{itemize}

The full discussion of the classification is given in Appendix \ref{app:redshifts}, and summarized in Table \ref{tab:set1}. 
In total, 15 GRBs have been classified with a reliable redshift association (dataset 'A') and 7 additional GRBs with a probable redshift association (dataset 'B' with 22 data-points in total). The 6 GRBs with an implausible association are not considered in this analysis. 
A bias might exist whereby we tend to favor redshift estimates from nearby GRBs, since these are in general brighter and the underlying galaxy can be identified more easily, but this uncertainty is not taken into account. 

%%%%%%%%%%%%%%%%%%%%%%%%%%%%%%%%%%
\section{Description of the fit model}

We describe the model used to fit the observed redshift values, and the parameters needed to convert the fit results into an astrophysical rate.

\subsection{The fit model} 

We now describe the model used for the fit, which is a general one of the rate of astronomical objects as a function of their cosmological redshifts. 
These models are described e.g. by \citet{Chapman:2007xs,Chapman:2008zx} and \citet{guetta:2005} and are used to model the distribution of type-I GRBs. 
This model expresses the number of observable objects with a redshift lower than some redshift $z_*$, and is given by: 

\begin{equation}
 N( z_*) = N_0 \int_0^{z_*} dz\;\frac{R(z)}{1+z} \; \frac{dV(z)}{dz} \; \int_{L_{min}(P_{lim},z)}^{L_{max}} \Phi(L) dL\;. \label{eq:generalFitfunction}
\end{equation}

In this equation, $N(z_*)$ is the number of type-I GRB above some minimum luminosity with a redshift lower than $z_*$, $R(z)$ is the rate-function (in units per volume) at a redshift $z$, $\Phi(L)$ is the luminosity function, and $dV(z)/dz$ is the volume of a comoving shell at redshift $z$.
The standard cosmology parameters used are $\Omega_M$=0.27, $\Omega_\lambda$=0.73, and h=0.71.
The rate function $R(z)$ describes the change in the intrinsic rate of objects as a function of redshift $z$, and all functions used for the fits are described in Appendix \ref{app:rates}. 
The luminosity function $\Phi(L)$ describes the distribution of sources as a function of their luminosity, which are given in Appendix \ref{app:luminosities}. 
The upper integration limit is set to $L_{max}=10^{55} \textrm{ergs}$, while the lower limit depends on the threshold of the satellite and the redshift as

\begin{equation}
L_\textrm{min}(z) = \frac{4\pi}{1+z}\, D_\textrm{lum}^2(z)\, P_\textrm{lim}\;.
\end{equation}

The detection threshold $P_{lim}$ is taken for the SWIFT satellite, and is roughly $P_{lim}\approx 10^{-8} \textrm{ergs}/\textrm{cm}^2$ as can be seen in Figure 12 of \citet{sakamoto:2008}. 

The fit of the model is performed by a least squares fitting function from the scipy module of python\footnote{http://www.scipy.org}, which uses a modified version of the Levenberg-Marquardt algorithm to minimize a function representing the difference between the observed data and the fitted model.
For each dataset, every possible combination of rate function and luminosity function is used in Eq. (\ref{eq:generalFitfunction}), yielding 16 different fit functions for each dataset. 
A  Kolmogorov-Smirnov (KS) probability is calculated to estimate the goodness-of-fit of each of these fit functions, and only fits with a KS probability of more than \minks~are kept for further analysis.

The norm of this equation, $N_0$, denotes the number density of GRBs at zero redshift. 
To obtain the local rate $r_\textrm{local}$, one need to take into account the observing period ($T=5$ years) and the fraction of type-I GRBs  used for the fit, compared to the total number of type-I GRBs observed during this period. The latter correction factors are $f=$28/15 for dataset 'A' and $f=$28/22 for dataset 'B'. 
The local rate is given as

\begin{equation}
r_\textrm{local} = \frac{f\;N_0}{T}\; .
\end{equation}

This yields the local, uncorrected rate, which has to be corrected for several effects, as described in the next subsection.

\subsection{Model-dependent parameters}

Several corrections must be applied to the local rate to obtain the true rate of binary mergers.  
These corrections include the beaming factor of GRBs $f_b^{-1}$, the field-of-view of the satellite $\upsilon$, the fraction of mergers producing a type-I GRB, $\eta$, and the fraction of type-I GRBs created by a merger, $\sigma$. 
The general expression to obtain the true, corrected rate from the uncorrected, local rate is

\begin{equation}
\label{eq:truerate0}
 r_\textrm{corr} = r_\textrm{local} \frac{f_b^{-1}\, \sigma}{\eta\; \upsilon}\;.
\end{equation}

\subsubsection{Beaming factor}

There is convincing evidence that the outflows of GRBs are strongly beamed \citep{1538-4357-522-1-L39,1538-4357-523-2-L121,Grupe:2006uc,Burrows:2006ar}, which we consider in this work. 
Since gamma-ray bursts are only visible if the Earth is inside the cone of the outflow, the true rate will be higher than deduced from the redshift fits alone, by a factor $f_b^{-1}=1/(1-\cos(\theta))$, the \textit{beaming factor}. 
The angle $\theta$ is the opening angle of the outflow.

\begin{table*}
\caption{List of opening angles or their lower limits. }
\centering
\begin{tabular}{lll} 
\hline 
\textbf{GRB} & \textbf{angle in degrees} & \textbf{beaming factor $f_b^{-1}$ with correction} \\ \hline
 050709 & $>$1.5\tablefootmark{a} & $<9000$ \\ \hline
 050724 &  8-15\tablefootmark{a}, $>$25\tablefootmark{b}  & $<30$ or $\sim$90-300 \\ \hline
 051221A & 4\tablefootmark{c}, 3.7\tablefootmark{d}  & $\sim$1250 or $\sim$1470 \\ \hline
 061021 & $4.5\zeta=4.8$\tablefootmark{d} & $\sim$870 \\ \hline
\end{tabular}
\tablefoot{The values in this table are obtained from the angle by using $f_b^{-1}=1/\left(1-\cos(\frac{4}{7}\theta)\right)$; see text for details, including the definition of $\zeta$.  References:
\tablefoottext{a}{\citet{Panaitescu:2006}};
\tablefoottext{b}{\citet{grupe-2006-653}};
\tablefoottext{c}{\citet{2006ApJ...653..468B}};
\tablefoottext{d}{\citet{Racusin:2008bx}}
}
\label{tab:opening} 
\end{table*}

In the standard model of GRBs (see e.g. \citet{RevModPhys.76.1143, meszaros:2006} for reviews), the outflowing matter, initially with a Lorentz factor of $\Gamma_0$, is confined to a cone of opening angle $\theta$. 
This angle changes when the Lorentz factor of the outflow decelerates and becomes comparable to $\theta^{-1}$; the jet then starts spreading sideways and an achromatic drop in the light curve (\textit{jet-break}) is expected. Calculations have been performed that  show a dependency of the jet-break $t_b$ on the jet opening angle as $\theta = 0.057 \,\zeta\, t_b^{3/8}$~\citep{1538-4357-519-1-L17,1538-4357-562-1-L55}, with $\zeta$ defined as

\begin{equation}
\label{eq:zeta}
\zeta = \left( \frac{1+z}{2}\right)^{-3/8} \left( \frac{\eta_\gamma}{0.2} \right)^{1/8} \left( \frac{E_{\gamma,\textrm{iso}}}{10^{53} \textrm{ergs}} \right)^{-1/8} \left( \frac{n}{0.1 \textrm{cm}^{-3}} \right)^{1/8} \,\,.
\end{equation}

Here $z$ is the redshift of the source, $\eta_\gamma$ the efficiency of converting the energy of the outflow into gamma rays, $n$ is the mean circumburst density, and $E_{\gamma,\textrm{iso}}$ the isotropic equivalent gamma-ray energy.

It should be noted that the observer is in general not directly located on the jet axis, but at an angle $\theta_\textrm{off}$ relative to this axis \citep{vanEerten:2010zh}, so that $\theta_\textrm{meas} = \theta_\textrm{true} +\theta_\textrm{off}$.
A typical observer is more likely to be positioned at a large angle off-axis, with a mean value of

\begin{equation}
 \langle \theta_\textrm{off} \rangle = \int_0^{\theta_\textrm{true}} p(\theta)\, \theta\, d\theta = \frac{3}{4}\theta_\textrm{true} \,\,,
\end{equation}

where we have used the normalized probability density function $p(\theta)=3\theta^2/\theta_\textrm{true}^3$.
The true average opening angle of a GRB is then given by 

\begin{equation}
 \theta_\textrm{true}= \theta_\textrm{meas}-\theta_\textrm{off} \simeq  \frac{4}{7}\theta_\textrm{meas}\, ,
\end{equation}

which is only about half the angle inferred by measurements of the jet-break times as explained above. \\

Evidence of jet-breaks in type-I GRBs are very rare, and summarized in Table \ref{tab:opening}. 
The following list gives some more details about these findings:

\begin{itemize}
\item \textbf{GRB~050709}: The afterglow of this GRB has was in \citet{Panaitescu:2006} where no jet-break was found within 10 days of the time of the burst. Two different circumburst densities have been used ($n=10~\textrm{cm}^{-3}$ and $n=10^{-5} \textrm{cm}^{-3}$), with the latter being more realistic in the case of a type-I GRB, yielding an opening angle of  $\theta>6^\circ$ \citep{Panaitescu:2006}.
\item \textbf{GRB~050724}: The afterglow for this GRB was also analyzed by \citet{Panaitescu:2006} who claimed  to have seen a jet-break one day after the burst time and suggested an opening angle of $\theta=10-15^\circ$ or a somewhat tighter value of  $\theta=8-12^\circ$, assuming energy injected into the afterglow from long-lived X-ray flares \citep{Panaitescu:2006}. The density was assumed to be in the range $n=0.1-1000\, \textrm{cm}^{-3}$.
This result is doubted by \citet{Grupe:2006uc} who claimed that the opening angle is larger than $25^\circ$ for $n=0.1\, \textrm{cm}^{-3}$.
\item \textbf{GRB~051221A}: The analysis of the lightcurve for this GRB identified three breaks, the last one of which was assumed to be the jet break, corresponding to an opening angle of $\theta=4^\circ$ or $\theta=8^\circ$, depending on the ambient density ($n=10^{-4}\, \textrm{cm}^{-3}$ and $n=0.1\, \textrm{cm}^{-3}$, respectively) \citep{Burrows:2006ar}. Again, the low-density value seems more appropriate for type-I GRBs. A value of $\theta=3.7^\circ$ is given in \citet{Racusin:2008bx}. 
\item \textbf{GRB~061021}: The jet-break times and the corresponding opening angles for this GRB can be found in \citet{Racusin:2008bx}, with an assumed circumstellar density of n$\simeq1\, $cm$^{-3}$. In the case of a type-I GRBs, however, the more probable place is a much less dense region, so a choice of  n$\simeq 10^{-4} $cm$^{-3}$ seems more appropriate. Using the mean redshift for type-I GRBs in equation  (\ref{eq:zeta}) and a mean isotropic equivalent energy of $E_{\gamma,ISO}\simeq 10^{50}$~erg, a more realistic value for $\zeta$=1.0573 has been used.
\end{itemize}

\begin{figure}
\includegraphics[width=3.5in,angle=0]{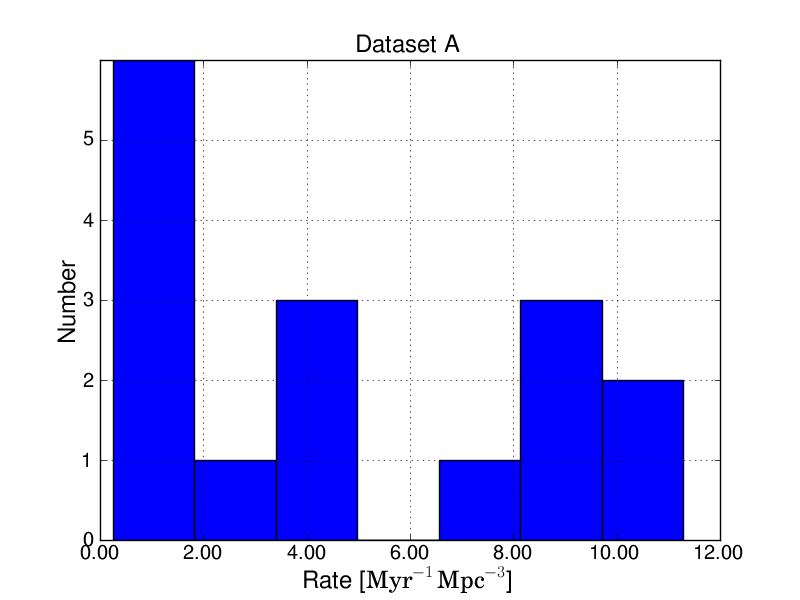}
\includegraphics[width=3.5in,angle=0]{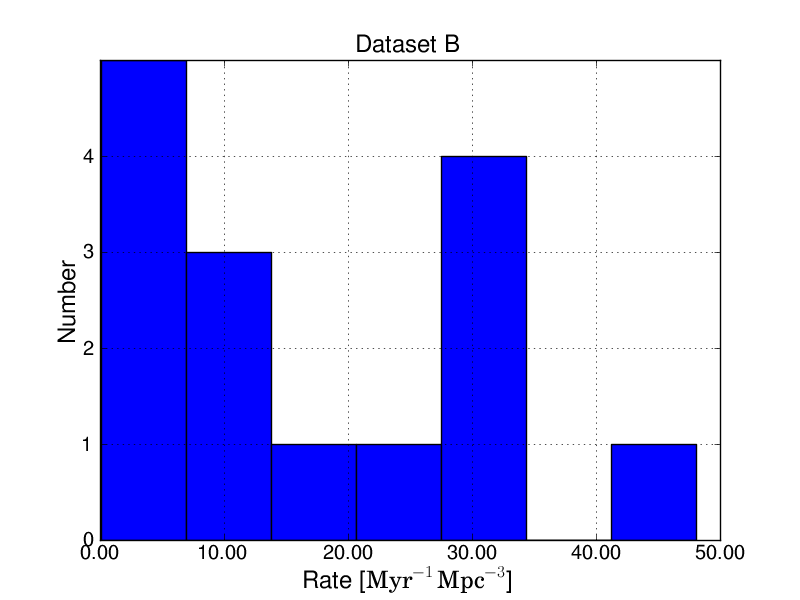}
\caption{Histogram of the fit results across the different rate functions and luminosity functions (as listed in Table \ref{tab:fitresult2}). The left plot shows the result for dataset 'A' (reliable redshift values), while the right plot shows the results for dataset 'B' (reliable and probable redshift values). The individual rates are always above zero with a minimum rate of 0.25\tunit for dataset 'A' and 0.066\tunit for dataset 'B'. }
\label{fig:histFit}
\end{figure}

\begin{figure}
\includegraphics[width=3.5in,angle=0]{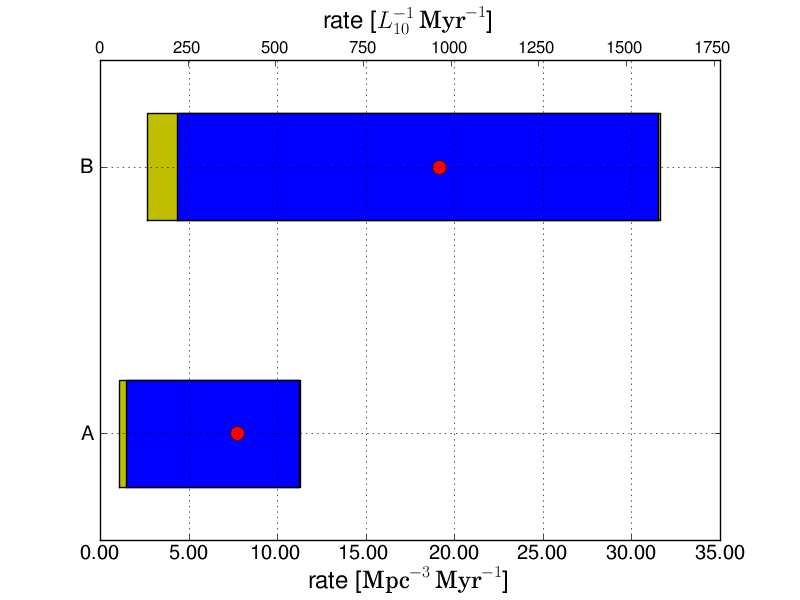}
\caption{Ranges of the computed rates using dataset 'A' with only the reliable redshift values and dataset 'B' with both the reliable and probable redshift values. The red dot indicates the median value, the blue bar the 20\% to 80\% quantile range, while the yellow bar covers the total rate range. }
\label{fig:barFit}
\end{figure}

The findings of the opening angles of type-I GRBs are summarized in Table \ref{tab:opening}, taking into account the correction factor of $4/7$. 
The beaming factors range between $<30$ to up to $\sim$9000, which represents either the observational spread in the real value or a multimodal distribution from different processes (i.e. SGR, NS-NS merger, NS-BH merger). 
In the remainder of this paper, a value of $f_b^{-1}=$500 is used, corresponding to an angle of 6.3$^\circ$, which seem to be a reasonable choice and agrees with the range of $13\lesssim f_b^{-1}\lesssim 10^4$ as found in \citet{0004-637X-638-2-930} and \citet{0004-637X-576-2-923}. 
We note that these estimates are based on only four reliable jet break measurements. 
This might imply that the majority of GRBs have jet breaks at a much later time, leading to a much larger beaming factor.
However, the four presented cases indicate that at least some GRBs have a beaming factor in the deduced range.\\

\subsubsection{Fraction of type-I GRB originating from a merger}

Not every type-I GRB is caused by a merger event, some possibly being created from a soft-gamma repeater or different process. 
As up to 25\% of all type-I GRBs might be created by a SGR \citep{Tanvir:2005,Levan:2008}, an assumption that 75\% have a merger progenitor seems reasonable; therefore a value of $\sigma=0.75$ will be assumed in the remainder of this paper.
Since this is the basic assumption on which this work is based, we consider this value to be fixed.

\subsubsection{Fraction of mergers producing a type-I GRB}

Not every merger will lead to a type-I GRB, because one material object is needed to create the observed ultra-relativistic outflow.
This implies that one of the two objects must be a neutron star, while the other object is either a neutron star or a black hole.
Even then, it is hard to estimate the fraction of mergers that might create a type-I GRB; a recent investigation inferred  $\eta=0.01-0.4$ in the case of NS-BH systems \citep{Belczynski:2007xg}, with an even larger spread in the case of NS-NS systems  \citep{Belczynski:2007wt}. 
The recent discovery of a 2~$M_\odot$ neutron star \citep{Demorest:2010nat,Demorest:2010bx} makes it far more plausible that two coalescing neutron stars generate type-I GRBs \citep{Ozel:2010bz}, resulting in a much larger value of $\eta$.
A value of $\eta=0.5$ is used as a first guess for the latter analysis, but we investigate also the outcome of choosing any possible value between 0 and 1.

\begin{figure}
%\epsscale{.80}
%\plotone{scatter_rate_minLOG.eps}
\includegraphics[width=4.0in,angle=0]{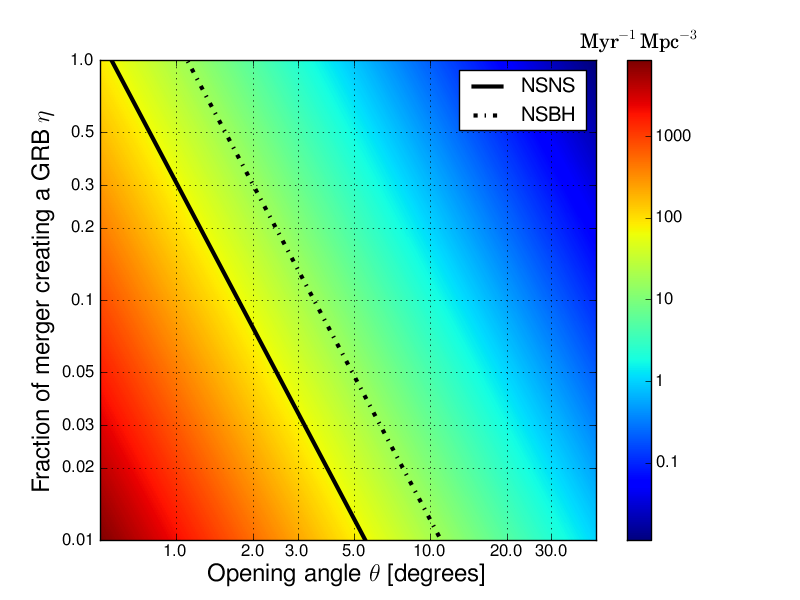}
\caption{Plot of the rate for model 'delay power', which yields the minimum rate (see Appendix \ref{app:rates}), as a function of the GRB opening angle $\theta$ and the parameter $\eta$. The other parameters are kept constant as they are more reliable. The color show the rate in units of \tunit, and the black lines delineate the areas excluded by GW upper limits in the case of NS-NS and NS-BH binaries \citep{LVC:2010yb}; the lower left areas are excluded.}
\label{fig:scatterrate}
\end{figure}

\subsubsection{Field-of-view}

Most of the GRBs considered in this work are detected by SWIFT, which has a field-of-view (FOV) of 1.4 sr half-coded \citep{Barthelmy:2005hs}. 
This corresponds to about 10\% visibility of the sky ($\upsilon=0.1$) at any time for SWIFT. 
Although GRB data from other missions have been used\footnote{from HETE, http://space.mit.edu/HETE/} and the actual sky coverage is not constant, a value of 10\% seems reasonable throughout the period considered in this paper. 
We also assume a 100\% duty cycle of the satellites over the entire period of five years we consider, which is not exactly true, but the error associated with this parameter is much smaller than the other parameters.

%%%%%%%%%%%%%%%%%%%%%%%%%%%%%%%%%%%%%%%%%%%%%%%%%%%%%%%%%%%%
\section{Fit results and discussion}

Following the above discussion of the parameters, the true corrected rate $r_\textrm{corr}$ can be obtained from the uncorrected local rate $r_\textrm{local}$ with the expression

\begin{equation}
\label{eq:truerate}
 r_\textrm{corr} = 7500\; r_\textrm{local}\,\left( \frac{f_b^{-1}}{500}\right)\,\left( \frac{\sigma}{0.75}\right) \, \left( \frac{0.50}{\eta}\right)\,\left(\frac{0.10}{\upsilon}\right)\;,
\end{equation}

in which the prefactor clearly reflects the default choices (and is equal to $500\cdot0.75/(0.50\cdot0.10)$).

Table \ref{tab:fitresult2} summarizes the final sample of models, including their fitted parameters and the goodness-of-fit values, when using the parameters shown in equation (\ref{eq:truerate}).
For dataset 'A', the lowest rate (1.1\tunit) is obtained from the model with the delay rate function and power luminosity function, while the maximum rate (11.3\tunit) is obtained from the model with a constant rate function and lognormal luminosity function. 
The histogram of the rates are shown, for both datasets, in Figure~\ref{fig:histFit}, and in Figure~\ref{fig:barFit} as a bar plot, indicating the 20\%, 50\% and 80\% quantiles. 
For dataset 'B', the rates are between 2.6 and 31.6~\tunit{}.
As dataset 'B' includes redshift values that are more uncertain and on average closer, it is unsurprising that this set yields higher rate values.
To be conservative, only results from dataset 'A' are used hereafter.

%%%%%%%%%%%%%%%%%%%%%%%%%%%%%%%%%%%%%%%%
\subsection{Constraints on GRB parameters}

The effect of the uncertainty in the parameters $f_b^{-1}$ and $\eta$ is shown in Fig. \ref{fig:scatterrate}, which shows the minimum rate obtained from a model (i.e. model 'delay power' in Table \ref{tab:fitresult2}), as a function of the opening angle $\theta$ and the fraction of mergers creating a type-I GRB, $\eta$. 
These lower limits vary between 0.01\tunit{} and $10^4$\tunit{}, while the upper limits are between 0.1\tunit{} and $10^5$\tunit{}. 
The black lines show the 90\% upper limits to merger rates as a result of  LIGO/Virgo searches, which are 43.6\tunit{} for NS-BH mergers and 172\tunit{} for NS-NS mergers \citep{LVC:2010yb}.
The excluded areas to the lower left of these lines impose constraints on the opening angle $\theta$ of the outflow in GRBs.
When assuming that most type-I GRBs are created by NS-BH mergers, an opening angle of smaller than $\sim1^\circ$ is excluded. 
This plot also indicates that $\eta$ might not be too small, agreeing with the results obtained in \citet{Belczynski:2007xg} and \citet{Ozel:2010bz}.

%%%%%%%%%%%%%%%%%%%%%%%%%%%%%%%%%%%%%%%%
\subsection{Comparison with other rate estimates}

\begin{table}[t!]
\caption{Comparison of realistic rate estimates from this work with the estimates given in \citet{LVC:2010cf}. }
\centering
\begin{tabular}{|l|l|l|l|l|}
\hline
  & $R_{low}$ & $R_{re}$ &  $R_{high}$ & $R_{max}$ \\  \hline
NS-NS (\tunit) & 0.01             & 1     & 10  & 50 \\
NS-BH (\tunit) & $6\times10^{-4}$ & 0.03  & 1   &  \\
this work (\tunit) & 0.03 & \ratere & --- & $10^4$ \\ \hline
\hline
  & $R_{low}$ & $R_{re}$ &  $R_{high}$ & $R_{max}$ \\  \hline
NS-NS ($L_{10}^{-1} \textrm{Myr}^{-1}$)     & 0.5   & 50  & 500 & 2500 \\
NS-BH ($L_{10}^{-1} \textrm{Myr}^{-1}$)     & 0.03  & 1.5   & 50  &  \\
this work ($L_{10}^{-1} \textrm{Myr}^{-1}$) & 1.5 & 390& ---& $5\times10^{5}$ \\ \hline
\end{tabular}
\tablefoot{In this table, $R_\textrm{low}$ is identified with the pessimistic estimate (the minimum rate obtained in this work), $R_\textrm{pl}$ with the plausible estimate (the median rate obtained in this work), $R_\textrm{high}$ with the plausible optimistic estimate, and $R_\textrm{max}$ with the upper limit (the upper limit in this work as well). The conversion factor between the two units is 0.0198 $L_{10}/\textrm{Mpc}^3$ \citep{Kopparapu:2007ib}. }
\label{tab:compareRates}
\end{table}

We now compare the rates deduced in this paper with other rate estimates. Two cases are considered: The coalescence rate of two neutron stars and the rate of coalescence of a neutron star with a black hole. 
For a NS-NS merger the rate is deduced from known binary pulsars in our Milky Way, and was estimated to be  realistically around 1\tunit, although they could be as high as 50\tunit{} \citep{LVC:2010cf}.
The rates predicted for NS-BH are much more uncertain, and have been estimated using population synthesis models. Realistic rates are around 0.03\tunit{}, although they could be as high as 1\tunit{} \citep{OShaughnessy:2008}.
A recent investigation of star formation found the rate of NS-BH binaries is possibly significantly higher by a factor of $\sim$20, when a lower metallicity is assumed in the models \citep{Belczynski:2010tb}.

The rates estimated in this work range from \ratemin\tunit{} to \ratemax\tunit, with a median value at $\sim$\ratere\tunit{} for the choice of plausible parameters used at the beginning of this section. 
When including the uncertainties in the beaming factor $f_b^{-1}$, ranging from 30 (corresponding to 26$^\circ$) to 9000 (corresponding to 1.5$^\circ$), and for the fraction of mergers producing a type-I GRB $\eta$, from 0.01 to 1.0, the minimum rate becomes 0.03\tunit{}, while the maximum rate becomes 10200\tunit{}. 
Table~\ref{tab:compareRates} compares these rate estimates with the ones given in  \citet{LVC:2010cf}.
The plausible pessimistic rate estimate $R_{low}$ is identified with the minimum rate in the present study, and the 
realistic rate estimate $R_{re}$ is identified with the median value estimated here using the plausible parameter choices. 
This value, $\sim$\ratere\tunit{}, is very similar to the high rate estimate $R_{high}$ for NS-NS mergers given in \citep{LVC:2010cf} (see Table~\ref{tab:compareRates}). 
It should also be noted that the maximum rate estimate of 10200\tunit{} is much higher than the excluded values from LIGO/Virgo measurements, which are 43.6\tunit{} for NS-BH mergers and 172\tunit{} for NS-NS mergers \citep{LVC:2010yb}.
For completeness, Table~\ref{tab:compareRates} also shows the rates in units of $L_{10}^{-1} \textrm{Myr}^{-1}$.

%%%%%%%%%%%%%%%%%%%%%%%%%%%%%%%%%%%%%%%%%%%%%%
\section{Summary and conclusion}

This work has utilized redshift measurements of type-I GRBs (i.e. short GRBs) to obtain the local rate of NS-NS and NS-BH mergers, respectively. 
The list of available redshifts have been revisited in detail, to assess the reliability of each redshift value. 
For the 22 type-I GRBs observed between 2004 and 2009, 15 have been found to have reliable redshifts, and 7 to have probable values; the redshifts of the remaining six type-I GRB were found to be too uncertain and were excluded from the analysis. 

A cumulative distribution was constructed using the two datasets of redshift values, which had been fitted by models using different functions for the rate and luminosity. 
A KS-test criterion was used to select models with reliable fits.
To obtain the true local rate, the fit results were corrected for several factors, including the fraction of mergers producing a type-I GRB, $\eta$, the number of type-I GRBs created by a merger, and the beaming factor of the GRBs. 

The obtained rates are consistent with the high-rate estimates given in \citet{LVC:2010cf}, with a median rate of \ratere \tunit. 
When including the uncertainties from the beaming factor and parameter $\eta$, the rate is found to vary between 0.01\tunit{} and $10^5$\tunit.
However, results from LIGO/Virgo observations, placing an upper limit on the rates of NS-NS and NS-BH mergers, can be used to constrain the opening angles of GRBs; the investigation indicates that opening angles of $\lesssim 1^\circ$ are excluded. 

Further work is required to improve the accuracy of the results, i.e. by taking into account the fluxes measured for GRBs, or constraining the ranges of some parameters used in this work. 
Ultimately, only direct detections of gravitational waves from mergers will yield a more precise rate, and if associated with type-I GRBs, the physical properties of gamma-ray bursts will also be able to be constrained.

%%%%%%%%%%%%%%%%
\begin{acknowledgements}
I would like to thank Chris Belczynski and Tomasz Bulik for providing many useful ideas and suggestion, and especially Frederique Marion and Benoit Mours for carefully reading this manuscript. 
I also like to thank the anonymous referee for useful comments, and the support from Astronomy and Astrophysics.
\end{acknowledgements}

%%%%%%%%%%%%%%%%%%%%%%%%%%%%%%%%%%%%%%%%%%%%%%%%
\appendix

%%%%%%%%%%%%%%%%%%%%%%%%%%%%%%%%%%%%%%%%%%%%%%%%
\section{Detailed redshift discussions}
\label{app:redshifts}

\paragraph{GRB~020531}

HETE detected this short GRB with a duration of 0.2-1 second [GCN1399]\footnote{Each citation from the Gamma-ray Coordinate Netwok (GCN) will be given in this format. The citation can be found on the website \texttt{http://gcn.gsfc.nasa.gov/gcn3\_archive.html}} and follow-up observations revealed various sources in the IPN error circle [GCN1408,GCN1415], including two new asteroids [GCN1400]. One of these sources was found to weaken in brightness [GCN1426] very close to an extended object, i.e. the probable host galaxy [GCN1427], which has a spectroscopic redshift of z=1.0 [GCN1428]. A different fading source was found just outside the error box [GCN1434], making this a probable redshift estimate only.

\paragraph{GRB~040924}

This is a short GRB with a duration of $\sim$1.5~s [GCN2754] for which an optical afterglow (OA) has been identified [GCN2734]. A spectroscopic analysis of the very likely host galaxy [GCN2750] revealed a redshift of z=0.859 [GCN2800].

\paragraph{GRB~050416}

This is a SWIFT/XRT GRB [GCN3264,GCN3268] with a duration of 2.4\ui0.2 seconds [GCN3273], and because an optical transient was observed [GCN3265,GCN3266] the measured redshift of z=0.6535$\pm$0.0002 [GCN3542] is very likely.

\paragraph{GRB~050509B}

GRB~050509B was found only within the XRT error circle [GCN3395] with a radius of 8''. Further investigations showed that there are at least four more sources in the XRT error circle, one of them a probable high-redshift galaxy [GCN3401]. The chance association of a low redshift galaxy is reported to be very small  [GCN3418], hence this redshift estimation is implausible.

\paragraph{GRB~050709}

Although a long duration is reported for this GRB of $\sim$130 seconds [GCN3653], the lightcurve and other spectral features classify it as a type-I GRB [GCN3570,GCN3653]. The association of the  afterglow with the probable host galaxy [GCN3605,GCN3612] makes the redshift estimation of z=0.16 reliable. 

\paragraph{GRB~050724}

GRB~050724 has a $T_{90}$ duration of strictly $3\pm1$ seconds, but because it could belong to the type-I GRB class [GCN3667] it is considered in the sample. Four objects were found in the XRT error circle [GCN3672], of which two are identified as Galactic stars [GCN3675,GCN3679]. Additional observations hale us improve our confidence, that the object labelled as ``D'' is the host galaxy with a redshift of 0.258 [GCN3690,GCN3700].

\paragraph{GRB~050813}

This SWIFT GRB with a duration of 0.6$\pm$0.1 seconds [GCN3793] has been found at a position with several faint extended objects, probably forming a galaxy cluster at high redshift, which makes this cluster the most likely source of that GRB [GCN3798]. Measurements of the redshift of the galaxies suggest a value of z=0.722 [GCN3801], although a redshift of z=0.65 also seems plausible [GCN3808]. The more conservative value of 0.722 will be used in this work as a probable value.

\paragraph{GRB~050906}
SWIFT detected this very short GRB ($T_{90}=0.128\pm0.016$s [GCN3935]) without finding the source with XRT, leaving only the BAT error circle to search for an afterglow [GCN3927,GCN3935]. Since this circle contains the massive star-forming galaxy IC 328 at z=0.031, a galaxy cluster at z=0.43, and field galaxies of unknown redshift, any value is unlikely.

\paragraph{GRB~051016B}

For this SWIFT GRB with a duration of 4$\pm$0.1~s [GCN4104], a probable optical afterglow was found [GCN4111], in a galaxy with redshift z=0.9364 [GCN4186]; this makes the redshift value reliable.

\paragraph{GRB~051221A}

This GRB had a duration of 1.4$\pm$0.2 seconds [GCN4365], and an optical afterglow was detected [GCN4375]. The measured redshift of z=0.5465 [GCN4384] is therefore reliable.

%%%%%%%%%%%%%%%%%%%%%%%%%%%%%%%%%%%%%%%%%%%%
\paragraph{GRB~060502B}

GRB~060502B was a very short GRB with a duration of 0.09\ui0.02 seconds [GCN5064], for which two sources were found in the XRT error circle [GCN5066,GCN5071]. One is assumed to be a star, while the other appears to be an extended object, whose reliable redshift is measured to be z=0.287 [GCN5238].

\paragraph{GRB~060505}

This GRB has a nominal $T_{90}$ duration of $4\pm1$seconds [GCN5142], and therefore not clearly assigned to either type-I or type-II. The position of the optical afterglow was found to be 4``.3 from  the galaxy 2dFGRS S173Z112, with a redshift of z=0.089 [GCN5123]. The distance in projection of this late-type galaxy was found to be 7 kpc [GCN5123]. No supernova was found to be associated with this GRB [GCN5161], suggesting that this might be either a merger-driven GRB or a GRB at a much larger distance. The redshift value is classified as probable.

\paragraph{GRB~060801}

This short GRB (duration of 0.5\ui0.1 s [GCN5381]) was found in the SWIFT/XRT instrument [GCN5378], which revealed four objects in its field [GCN5384,GCN6386]. In the revised XRT error circle [GCN5389], two objects remained, of which one is extended. The redshift of that extended object is z=1.131 [GCN5470], making it a reliable estimation.

\paragraph{GRB~061006}

An optical afterglow was found for this GRB [GCN5718], revealing the reliable redshift to be z=0.4377\ui0.0002 \citep{berger2006_GRB061006}. The formal duration  is 130\ui10 s [GCN5704], but initial short spikes lasting $\sim$ 0.5 seconds, on which SWIFT did not trigger [GCN5702,GCN5710], places this GRB in the type-I category.

\paragraph{GRB~061201}

This short GRB (duration of $0.8\pm0.1$~seconds [GCN5882]) had an optical afterglow [GCN5896], but no galaxy was found at its position. Close-by objects include a galaxy at redshift 0.111 [GCN5952], as well as the galaxy cluster Abell 995, for which a mean redshift of z$\sim$0.0835 was determined [GCN5995]. The offset of the GRB from the galaxy would be 34 kpc in the first case, and 800 kpc in the second case (from the center of the cluster). 
The value of z=0.111 is used in the further analysis.\footnote{Using the value of 0.0835 changes the outcomes of the fits and the results of this work insignificantly - it is therefore safe to use a redshift value of 0.111.}

\paragraph{GRB~061210}

This GRB has a nominal duration of $T_{90}=85 \pm 5\;$s~[GCN5905], but an initial short spike of duration $\sim 60\;$ ms places it into the type-I regime. This GRB has been located in the XRT error circle containing three galaxies [GCN5922]. Since no optical transient was found, the association with a given galaxy as host is doubtful, which makes the redshift implausible.

\paragraph{GRB~061217}

This short GRB, with a $T_{90}$ of $0.30\pm0.05$ seconds [GCN5930], exhibited no optical afterglow, but the XRT position was found to be within 11 arcsec of a galaxy [GCN5948]. Two objects have been found in a more deeper observation, with the brighter object proposed to be the host galaxy of this GRB [GCN5949,GCN5953]. A subsequent observation of the proposed host galaxy yields a redshift of z=0.827 [GCN5965], so the source would have a isotropic-equivalent energy release of about $8\times 10^{49}$ erg [GCN5965], which is rather high for a type-I GRB. This makes the redshift probable only. 

%%%%%%%%%%%%%%%%%%%%%%%%%%%%%%%%%%%%%%%%%%%%
\paragraph{GRB~070209}

The localization of this GRB within the BAT error circle contained no single source within the BAT error circle, but three X-ray sources in its proximity [GCN6095]. None of these sources were found to be the afterglow of this GRB [GCN6119], making the measured redshift of the source closest to the GRB position of z=0.314 [GCN6101] very unlikely. 

\paragraph{GRB~070406}

A short GRB with a $T_{90}$ of $0.7\pm0.2$ seconds [GCN6261], for which a bright source was detected in the XRT error circle, whose redshift is 0.703 [GCN6262]. Spectral features indicate this to be a quasar, and unrelated to the burst. Further investigations showed no hint of a fading afterglow, and the large number of faint sources found in the XRT error circle make this redshift estimate of z=0.11 [GCN6249] uncertain. 

\paragraph{GRB~070429B}

This BAT GRB has a duration of $0.5\pm0.1$ s [GCN6365], and two faint objects were found in the XRT error circle [GCN6372]. For the brighter objects, a redshift of z=0.9023\ui0.0002 was determined [GCN7104]~\citep{Cenko:2008vt}, as well as evidence that this object contains a fading source [GCN7145].
Therefore, the measured redshift is reliable. 

\paragraph{GRB~070714B}

This long GRB with a standard $T_{90}$ time of 64\ui5 seconds shows spectral features of a type-I GRB, especially the zero spectral lag [GCN6623]. An optical transient was found in the XRT  [GCN6630], and the host's galaxy redshift is found to be z=0.9225\ui0.0001 [GCN6836]\citep{Cenko:2008vt}, which makes this a reliable redshift.

\paragraph{GRB~070724}

This SWIFT GRB [GCN6654] had a duration of 0.4$\pm$0.04 seconds [GCN6656], and two sources were found in the XRT error circle, none of which showed variations [GCN6673]. The redshift of one of the source was found to be z=0.457 [GCN6665], making this a probable estimate.

\paragraph{GRB~070810B}

This is a short GRB with a duration of only 0.08$\pm$0.01 seconds [GCN6753]. Several possible sources has been identified in the XRT error circle, among them a nearby bright galaxy at z=0.0385 (source S1) and a cluster of galaxies at a redshift of z=0.49 [GCN6756] with an X-ray source (source S2 in [GCN6754]). In a latter observation, the second source no longer was detected, making this the probable position of the afterglow with a redshift of 0.49.

\paragraph{GRB~071227}

For this GRB, an optical afterglow was found [GCN7157] to coincide with the single source in the XRT error circle [GCN7151]. The redshift of z=0.384 [GCN7152,GCN7154] is therefore reliable.

%%%%%%%%%%%%%%%%%%%%%%%%%%%%%%%%%%%%%%%%%%%%%%%%%%%%%
\paragraph{GRB~080121}

The only redshift reported for this GRB is z=0.046 for two galaxies in the BAT error circle [GCN7210]. In this field, many other galaxies are present that might belong to a group of galaxies [GCN7210]. Since no XRT position could be determined [GCN7209], the redshift value is probable. 

\paragraph{GRB~080520}

For this short GRB with a duration of 2.8$\pm$0.7 seconds [GCN7761], an optical afterglow was found [GCN7753] for which a redshift of z=1.545 was determined [GCN7757]. This is a reliable redshift estimation.

%%%%%%%%%%%%%%%%%%%%%%%%%%%%%%%%%%%%%%%%%%%%%%%%%%%%%%%
\paragraph{GRB~090417A}

This GRB with a duration of 0.072$\pm$0.018 seconds [GCN9138], was found to be close to a low-redshift galaxy [GCN9134] with a redshift of 0.088 [GCN9136]. Since no optical afterglow has been found for this GRB, the redshift values are implausible. 

\paragraph{GRB~090510}

For this short GRB with a duration of 0.3$\pm$0.1 seconds [GCN9337] an optical afterglow has been found [GCN9338], which gives a reliable redshift of z=0.903 [GCN9353].

%%%%%%%%%%%%%%%%%%%%%%%%%%%%%%%%%%%%%%%%%%%%%%%%
\section{Rate models}
\label{app:rates}

The rate functions used to fit the data according to Eq. (\ref{eq:generalFitfunction}), except for the trivial case of the '\textbf{constant}' rate, are given by:

\begin{enumerate}

\item The '\textbf{sfr}' rate that follows the star formation rate (model 'SF2') described in \citet{porciani:2001} and \citet{guetta:2005}:
\begin{equation}
 R(z)\equiv R_{SF2}(z)= R_{s,0}\frac{23 \exp{(3.4\,z)}}{\exp{(3.4\,z)}+22} \;.
\end{equation}

\item The '\textbf{merger}' rate of two compact objects was derived in \citet{guetta:2005} by studying six observed double neutron stars \citep{champion:2004}. This rate follows a time-delay distribution ($\propto 1/\tau$)

\begin{equation}
 R(z)=R_{M,0} \int_0^{t(z)} d\tau \,R_{SF2}\left(z(t-\tau)\right)/\tau  \; .
\end{equation}

\item The '\textbf{delay}' rate similar to the 'merger' rate, but with a \textit{constant} time-delay distribution

\begin{equation}
 R(z)  = R_{D,0} \int_0^{t(z)}  d\tau\; R_{SF2}\left(z(t-\tau)\right)  \;.
\end{equation}

\end{enumerate}

%%%%%%%%%%%%%%%%%%%%%%%%%%%%%%%%%%%%%%%%%%%%%%%%

\section{Luminosity models}
\label{app:luminosities}

This section describes the luminosity functions used to fit the data according to Eq. (\ref{eq:generalFitfunction}).

\begin{enumerate}
 \item A \textbf{single power-law} distribution, which is often used to describe the pdf of luminosity in astrophysics (with two parameters: $\Phi_0$ and $\alpha$)

\begin{equation}
 \Phi(L)=\Phi_0\, \left( \frac{L}{L_0}\right)^{-\alpha}
\end{equation}

 \item A \textbf{broken power-law} distribution, describing e.g. two underlying populations in the luminosity \citep{guetta:2005} (with four parameters: $\Phi_0$, $L_0$, $\alpha$ and $\beta$)

\begin{eqnarray}
\label{eq:broken}
 \Phi(L)= &\Phi_0\,  \left( \frac{L}{L_0}\right)^{-\alpha} &\quad \textrm{for}\; L<L_0 \;, \\
 \Phi(L)= &\Phi_0\,  \left( \frac{L}{L_0}\right)^{-\beta}  &\quad \textrm{for}\; L>=L_0 \;.
\end{eqnarray}

\item The \textbf{Schechter distribution}, used for example in \citet{Andreon:2006} (with three parameters: $\Phi_0$, $L_0$ and $\alpha$)
\begin{equation}
 \Phi(L)=\Phi_0\, \left( \frac{L}{L_0}\right)^{-\alpha}  \exp(-L/L_0) \;.
\end{equation}

\item The \textbf{log-normal distribution}, describing a standard candle, e.g. a population with about the same luminosity (following \citep{Chapman:2008zx}, with three parameters: $\Phi_0$, $L_0$, and $\sigma$)

\begin{equation}
 \Phi(L)=\Phi_0\, \frac{1}{L}\,\exp\left(\frac{-(\log{L}-\log{L_0})^2}{2\sigma^2}\right) \; .
\end{equation}

\end{enumerate}

%%%%%%%%%%%%%%%%%%%%%%%%%%%%%%%%%%%%%%%%%%%%%%%%
\newpage
\section{Fit results}

%\begin{minipage}
\begin{table}[!h!] 
\begin{footnotesize}
\caption{Fit results with a KS-probability of at least~\minks, which have been used in this work. }
\label{tab:fitresult2}
\centering
 \begin{tabular}{|l|l|l|l|l|l|}
 \hline
Data & Model & $\ln{L_0}$ & $\alpha, \sigma$ & KS & Rate \\
&&&&&[$\textrm{Mpc}^{-3} \textrm{Myr}^{-1}$]\\ \hline \hline
A & constant power & --- & 1.8 & 0.99 & 9.22 \\ \hline
A & constant schechter & 50.5 & 1.8 & 1.00 & 11.24 \\ \hline
A & constant lognormal & 24.1 & 8.1 & 1.00 & 11.29 \\ \hline
A & sfr power & --- & 2.0 & 0.81 & 7.76 \\ \hline
A & sfr schechter & 49.3 & 1.9 & 1.00 & 9.20 \\ \hline
A & sfr lognormal & 45.3 & 2.5 & 1.00 & 8.59 \\ \hline
A & merger power & --- & 2.2 & 0.81 & 2.94 \\ \hline
A & merger schechter & 49.4 & 2.2 & 1.00 & 3.99 \\ \hline
A & merger lognormal & 44.0 & 2.7 & 1.00 & 3.74 \\ \hline
A & delay power & --- & 2.3 & 0.95 & 1.07 \\ \hline
A & delay schechter & 49.8 & 2.3 & 1.00 & 1.49 \\ \hline
A & delay lognormal & 35.5 & 4.6 & 1.00 & 1.47 \\ \hline
\hline 
B & constant power & --- & 1.9 & 0.99 & 23.91 \\ \hline
B & constant schechter & 55.1 & 1.9 & 0.96 & 31.28 \\ \hline
B & constant lognormal & -510.2 & 37.0 & 0.96 & 31.25 \\ \hline
B & sfr power & --- & 2.1 & 0.94 & 19.12 \\ \hline
B & sfr schechter & 54.4 & 2.2 & 1.00 & 31.60 \\ \hline
B & sfr lognormal & -383.5 & 28.7 & 1.00 & 31.51 \\ \hline
B & merger power & --- & 2.3 & 0.94 & 7.04 \\ \hline
B & merger schechter & 55.3 & 2.4 & 1.00 & 13.30 \\ \hline
B & merger lognormal & -317.1 & 24.2 & 1.00 & 13.27 \\ \hline
B & delay power & --- & 2.3 & 1.00 & 2.63 \\ \hline
B & delay schechter & 55.7 & 2.5 & 0.98 & 4.37 \\ \hline
B & delay lognormal & -316.0 & 23.9 & 0.98 & 4.35 \\ \hline

\end{tabular}
\end{footnotesize}
\end{table}
%\end{minipage}

%%%%%%%%%%%%%%%%%%%%%%%%%%%%%%%%%%%%%%%%

%\bibliographystyle{haa}
%\bibliography{/virgo/users/dietz/Documents/MyWork/Papers/Bibliography/allGCNbib,/virgo/users/dietz/Documents/MyWork/Papers/Bibliography/bib,/virgo/users/dietz/Documents/MyWork/Papers/Bibliography/iulpapers,/virgo/users/dietz/Documents/MyWork/Papers/Bibliography/bibold}

\clearpage
\begin{thebibliography}{63}
\expandafter\ifx\csname natexlab\endcsname\relax\def\natexlab#1{#1}\fi

\bibitem[{Abadie {et~al.}(2010)}]{S5GRB}
Abadie, J. {et~al.} 2010, ApJ, 715, 1453

\bibitem[{Abbott {et~al.}(2004{\natexlab{a}})}]{LIGO03}
Abbott, B. {et~al.} 2004{\natexlab{a}}, Phys. Rev. D, 69, 122001

\bibitem[{Abbott {et~al.}(2004{\natexlab{b}})}]{LIGOS1instpaper}
Abbott, B. {et~al.} 2004{\natexlab{b}}, Nucl. Instrum. Methods, A517, 154

\bibitem[{Abbott {et~al.}(2005{\natexlab{a}})}]{LIGOS2iul}
Abbott, B. {et~al.} 2005{\natexlab{a}}, Phys.~Rev.~D, 72, 082001

\bibitem[{Abbott {et~al.}(2005{\natexlab{b}})}]{LIGOS2macho}
Abbott, B. {et~al.} 2005{\natexlab{b}}, Phys.~Rev.~D, 72, 082002

\bibitem[{Abbott {et~al.}(2006)}]{LIGOS2bbh}
Abbott, B. {et~al.} 2006, Phys.~Rev.~D, 73, 062001

\bibitem[{Abbott {et~al.}(2007)}]{DAbbott:2007c}
Abbott, B. {et~al.} 2007, Phys. Rev. D, 77, 062002

\bibitem[{Abbott {et~al.}(2008)}]{grb070201}
Abbott, B. {et~al.} 2008, ApJ, 681, 1419

\bibitem[{Acernese {et~al.}(2006)}]{0264-9381-23-19-S01}
Acernese, F. {et~al.} 2006, Classical and Quantum Gravity, 23, S635

\bibitem[{Andreon {et~al.}(2006)Andreon, Cuillandre, Puddu, \&
  Mellier}]{Andreon:2006}
Andreon, S., Cuillandre, J.-C., Puddu, E., \& Mellier, Y. 2006,
  MNRAS, 372, 60

\bibitem[{Barish \& Weiss(1999)}]{Barish:1999}
Barish, B.~C. \& Weiss, R. 1999, Phys.\ Today, 52 (Oct), 44

\bibitem[{Barthelmy {et~al.}(2005{\natexlab{a}})}]{swift2}
Barthelmy, S.~D. {et~al.} 2005{\natexlab{a}}, Nature, 438, 994

\bibitem[{Barthelmy {et~al.}(2005{\natexlab{b}})}]{Barthelmy:2005hs}
Barthelmy, S.~D. {et~al.} 2005{\natexlab{b}}, Space Sci. Rev., 120, 143

\bibitem[{Belczynski {et~al.}(2010)Belczynski, Dominik, Bulik, O’Shaughnessy,
  Fryer, \& Holz}]{Belczynski:2010tb}
Belczynski, K., Dominik, M., Bulik, T., {et~al.} 2010, ApJ, 715, L138

\bibitem[{Belczynski {et~al.}(2008)Belczynski, O’Shaughnessy, Kalogera,
  Rasio, Taam, \& Bulik}]{Belczynski:2007wt}
Belczynski, K., O’Shaughnessy, R., Kalogera, V., {et~al.} 2008,  ApJ, 680, L129

\bibitem[{Belczynski {et~al.}(2007)Belczynski, Taam, Rantsiou, \& van~der
  Sluys}]{Belczynski:2007xg}
Belczynski, K., Taam, R.~E., Rantsiou, E., \& van~der Sluys, M. 2007, arXiv:astro-ph/0703131

\bibitem[{Berger {et~al.}(2007)Berger, Fox, Price, Nakar, Gal-Yam, Holz,
  Schmidt, Cucchiara, Cenko, Kulkarni, Soderberg, Frail, Penprase, Rau, Ofek,
  Burnell, Cameron, Cowie, Dopita, Hook, Peterson, Podsiadlowski, Roth,
  Rutledge, Sheppard, , \& Songaila}]{berger2006_GRB061006}
Berger, E., Fox, D.~B., Price, P.~A., {et~al.} 2007, ApJ,
  664, 1000

\bibitem[{Burrows {et~al.}(2006{\natexlab{a}})Burrows, Grupe, Capalbi,
  Panaitescu, Patel, Kouveliotou, Zhang, Meszaros, Chincarini, Gehrels, , \&
  Wijers}]{2006ApJ...653..468B}
Burrows, D.~N., Grupe, D., Capalbi, M., {et~al.} 2006{\natexlab{a}}, ApJ, 653,
  468

\bibitem[{Burrows {et~al.}(2006{\natexlab{b}})}]{Burrows:2006ar}
Burrows, D.~N. {et~al.} 2006{\natexlab{b}}, ApJ, 653, 468

\bibitem[{Campana {et~al.}(2006)}]{Campana:2006}
Campana, S. {et~al.} 2006, Nature, 442, 1008

\bibitem[{Cenko {et~al.}(2008)}]{Cenko:2008vt}
Cenko, S.~B. {et~al.} 2008, arXiv:0802.0874 [astro-ph]

\bibitem[{Champion {et~al.}(2004)}]{champion:2004}
Champion, D.~J. {et~al.} 2004, MNRAS, 350, L61

\bibitem[{Chapman {et~al.}(2008{\natexlab{a}})Chapman, Priddey, \&
  Tanvir}]{Chapman:2008zx}
Chapman, R., Priddey, R.~S., \& Tanvir, N.~R. 2008{\natexlab{a}}, arXiv:0802.0008 [astro-ph]

\bibitem[{Chapman {et~al.}(2008{\natexlab{b}})Chapman, Priddey, \&
  Tanvir}]{Chapman:2007xs}
Chapman, R., Priddey, R.~S., \& Tanvir, N.~R. 2008{\natexlab{b}}, AIP Conf.
  Proc., 983, 304

\bibitem[{Czerny {et~al.}(2011)Czerny, Janiuk, Cline, \&
  Otwinowski}]{Czerny:2010bd}
Czerny, B., Janiuk, A., Cline, D.~B., \& Otwinowski, S. 2011, New Astron., 16,
  33

\bibitem[{Demorest {et~al.}(2010{\natexlab{a}})Demorest, Pennucci, Ransom,
  Roberts, \& Hessels}]{Demorest:2010bx}
Demorest, P., Pennucci, T., Ransom, S., Roberts, M., \& Hessels, J.
  2010{\natexlab{a}}, arXiv:1010.5788 [astro-ph.HE]

\bibitem[{Demorest {et~al.}(2010{\natexlab{b}})Demorest, Pennucci, Ransom,
  Roberts, \& Hessels}]{Demorest:2010nat}
Demorest, P.~B., Pennucci, T., Ransom, S.~M., Roberts, M. S.~E., \& Hessels, J.
  W.~T. 2010{\natexlab{b}}, Nature, 467, 1081

\bibitem[{Eichler {et~al.}(1989)Eichler, Livio, Piran, \&
  Schramm}]{eichler:1989}
Eichler, D., Livio, M., Piran, T., \& Schramm, D. 1989, Nature, 340, 126

\bibitem[{Frail {et~al.}(2001)Frail, Kulkarni, Sari, Djorgovski, Bloom, Galama,
  Reichart, Berger, Harrison, Price, Yost, Diercks, Goodrich, \&
  Chaffee}]{1538-4357-562-1-L55}
Frail, D.~A., Kulkarni, S.~R., Sari, R., {et~al.} 2001, ApJ, 562, L55

\bibitem[{Fruchter(2006)}]{Fruchter:2006}
Fruchter, A.~S. 2006, Nature, 441, 463

\bibitem[{Gehrels {et~al.}(2005)}]{swift1}
Gehrels, N. {et~al.} 2005, Nature, 437, 851

\bibitem[{Grupe {et~al.}(2006{\natexlab{a}})Grupe, Burrows, Patel, Kouveliotou,
  Zhang, Meszaros, Wijers, \& Gehrels}]{grupe-2006-653}
Grupe, D., Burrows, D.~N., Patel, S.~K., {et~al.} 2006{\natexlab{a}}, ApJ, 653, 462

\bibitem[{Grupe {et~al.}(2006{\natexlab{b}})}]{Grupe:2006uc}
Grupe, D. {et~al.} 2006{\natexlab{b}}, ApJ, 653, 462

\bibitem[{Guetta \& Piran(2005)}]{guetta:2005}
Guetta, D. \& Piran, T. 2005, arXiv:astro-ph/0511239v2

\bibitem[{Harrison {et~al.}(1999)Harrison, Bloom, Frail, Sari, Kulkarni,
  Djorgovski, Axelrod, Mould, Schmidt, Wieringa, Wark, Subrahmanyan, McConnell,
  McCarthy, Schaefer, McMahon, Markze, Firth, Soffitta, \&
  Amati}]{1538-4357-523-2-L121}
Harrison, F.~A., Bloom, J.~S., Frail, D.~A., {et~al.} 1999, ApJ, 523, L121

\bibitem[{Hjorth {et~al.}(2003)}]{Hjorth:2003}
Hjorth, J. {et~al.} 2003, Nature, 423, 847

\bibitem[{Horvath(2002)}]{Horvath:2002}
Horvath, I. 2002, A\&A, 392, 791

\bibitem[{Kopparapu {et~al.}(2008)}]{Kopparapu:2007ib}
Kopparapu, R.~K. {et~al.} 2008, ApJ, 675, 1459

\bibitem[{Kouveliotou {et~al.}(1993)Kouveliotou, Meegan, Fishman, Bhat, Briggs,
  Koshut, Paciesas, \& Pendleton}]{Kouveliotou:1993}
Kouveliotou, C., Meegan, C.~A., Fishman, G.~J., {et~al.} 1993, ApJ, 413, L101

\bibitem[{Lee {et~al.}(2004)Lee, Ramirez-Ruiz, \& Page}]{Lee:2004xi}
Lee, W.~H., Ramirez-Ruiz, E., \& Page, D. 2004, ApJ, 608, L5

\bibitem[{Levan {et~al.}(2008)}]{Levan:2008}
Levan, A.~J. {et~al.} 2008, MNRAS, 384, 541

\bibitem[{Levinson {et~al.}(2002)Levinson, Ofek, Waxman, \&
  Gal-Yam}]{0004-637X-576-2-923}
Levinson, A., Ofek, E.~O., Waxman, E., \& Gal-Yam, A. 2002, ApJ, 576, 923

\bibitem[{{LIGO Scientific Collaboration and Virgo
  Collaboration}(2010{\natexlab{a}})}]{LVC:2010cf}
{LIGO Scientific Collaboration and Virgo Collaboration}. 2010{\natexlab{a}},
  Class. Quant. Grav., 27, 173001

\bibitem[{{LIGO Scientific Collaboration and Virgo
  Collaboration}(2010{\natexlab{b}})}]{LVC:2010yb}
{LIGO Scientific Collaboration and Virgo Collaboration}. 2010{\natexlab{b}},
  arXiv:1005.4655 [gr-qc]

\bibitem[{Mereghetti(2008)}]{Mereghetti:2008je}
Mereghetti, S. 2008, Astron. Astrophys. Rev., 15, 225, 0804.025

\bibitem[{M\'{e}sz\'{a}ros(2006)}]{meszaros:2006}
M\'{e}sz\'{a}ros, P. 2006, Rep. Prog. Phys., 69, 2259

\bibitem[{Narayan {et~al.}(1992)Narayan, Paczynski, \& Piran}]{narayan:1992}
Narayan, R., Paczynski, \& Piran, T. 1992, ApJ, 395, L83

\bibitem[{O'Shaughnessy {et~al.}(2008{\natexlab{a}})O'Shaughnessy, Kim,
  Kalogera, \& Belczynski}]{OShaughnessy:2008}
O'Shaughnessy, R., Kim, C., Kalogera, V., \& Belczynski, K. 2008{\natexlab{a}},
  ApJ, 672, 470

\bibitem[{O'Shaughnessy {et~al.}(2008{\natexlab{b}})O'Shaughnessy, Kalogera, \&
  Belczynski}]{O'Shaughnessy:2007fb}
O'Shaughnessy, R.~W., Kalogera, V., \& Belczynski, K. 2008{\natexlab{b}},
  ApJ, 675, 566

\bibitem[{Ozel {et~al.}(2010)Ozel, Psaltis, Ransom, Demorest, \&
  Alford}]{Ozel:2010bz}
Ozel, F., Psaltis, D., Ransom, S., Demorest, P., \& Alford, M. 2010, arXiv:1010.5790 [astro-ph.HE]

\bibitem[{Panaitescu(2006)}]{Panaitescu:2006}
Panaitescu, A. 2006, MNRAS, 367, L42

\bibitem[{Piran(2005)}]{RevModPhys.76.1143}
Piran, T. 2005, Rev. Mod. Phys., 76, 1143

\bibitem[{Porciani \& Madau(2001)}]{porciani:2001}
Porciani, C. \& Madau, P. 2001, ApJ, 548, 522

\bibitem[{Racusin {et~al.}(2009)}]{Racusin:2008bx}
Racusin, J.~L. {et~al.} 2009, ApJ, 698, 43

\bibitem[{Sakamoto {et~al.}(2007)}]{Sakamoto:2007}
Sakamoto, T. {et~al.} 2007, arXiv:0707.4626v3 [astro-ph]

\bibitem[{Sakamoto {et~al.}(2008)}]{sakamoto:2008}
Sakamoto, T. {et~al.} 2008, AJ Suppl. Series, 175, 179

\bibitem[{Sari {et~al.}(1999)Sari, Piran, \& Halpern}]{1538-4357-519-1-L17}
Sari, R., Piran, T., \& Halpern, J.~P. 1999, ApJ,
  519, L17

\bibitem[{Soderberg {et~al.}(2006)Soderberg, Nakar, Berger, \&
  Kulkarni}]{0004-637X-638-2-930}
Soderberg, A.~M., Nakar, E., Berger, E., \& Kulkarni, S.~R. 2006, ApJ, 638, 930

\bibitem[{Stanek {et~al.}(1999)Stanek, Garnavich, Kaluzny, Pych, \&
  Thompson}]{1538-4357-522-1-L39}
Stanek, K.~Z., Garnavich, P.~M., Kaluzny, J., Pych, W., \& Thompson, I. 1999,
  ApJ, 522, L39

\bibitem[{Tanvir {et~al.}(2005)}]{Tanvir:2005}
Tanvir, N.~R. {et~al.} 2005, Nature, 438, 991

\bibitem[{van Eerten {et~al.}(2010)van Eerten, Zhang, \&
  MacFadyen}]{vanEerten:2010zh}
van Eerten, H., Zhang, W., \& MacFadyen, A. 2010, arXiv:1006.5125 [astro-ph.HE]

\bibitem[{Woods \& Thompson(2004)}]{woods:2004}
Woods, P.~M. \& Thompson, C. 2004, in Compact Stellar X-Ray Sources, ed.
  W.~G.~H. Lewin \& M.~van~der Klis (Cambridge Univ. Press, Cambridge)

\bibitem[{Woosley \& Bloom(2006)}]{Woosley:2006}
Woosley, S.~E. \& Bloom, J.~S. 2006, ARA\&A, 44, 507

\bibitem[{Zhang {et~al.}(2009)}]{0004-637X-703-2-1696}
Zhang, B. {et~al.} 2009, ApJ, 703, 1696

\end{thebibliography}

\end{document}